\documentclass [a4paper,12pt, twoside] {article}
\usepackage{inputenc}
\usepackage[english]{babel}
\usepackage[dvips]{graphicx}
\usepackage{epsfig, latexsym, cite}

 \usepackage{amsfonts,amsmath,amssymb,amscd,array, euscript, eufrak}

 \newcommand {\bgc} {\begin{center}}
 \newcommand {\ec} {\end{center}}
 \newcommand {\sms} {\smallskip}
 \newcommand {\mes} {\medskip}
 
 \newcommand {\etit}[1] {\bgc{\textbf{\Large{#1}}} \mes\sms\ec}  
 \newcommand {\eauthm}[3] {\bgc{\textbf{\large{#1\\}} \mes\mes \emph{#2\\[3pt] #3}} \mes\sms\ec}
 
 \newcommand {\q} {\quad}
 \newcommand {\qq} {\qquad}

 \makeatletter

\renewcommand{\section}{\@startsection{section}{1}%
{\parindent}{3.6ex plus 0.8ex minus .2 ex} {1.6 ex plus .2 ex}{\large\bf}}   
 \makeatother

\begin{document}

\etit{Construction of the pseudo Riemannian geometry\\
 \mes on the base of the Berwald-Moor geometry}

\eauthm{D.G. Pavlov} {Bauman State Technical University, Moscow} {geom2004@mail.ru}

\eauthm{G.I. Garas'ko}{Russian Electrotechnical Institute, Moscow}{gri9z@mail.ru}

{\small The space of the associative commutative hyper complex numbers, $H_4$, is a
4-dimensional metric Finsler space with the Berwald-Moor metric. It provides the
possibility to construct the tensor fields on the base of the analytical functions
of the $H_4$ variable and also in case when this analyticity is broken. Here we
suggest a way to construct the metric tensor of a 4-dimensional pseudo Riemannian
space (space-time) using as a base the 4-contravariant tensor of the tangent
indicatrix equation of the Berwald-Moor space and the World function. The
Berwald-Moor space appears to be closely related to the Minkowski space. The break
of the analyticity of the World function leads to the non-trivial curving of the
4-dimensional space-time and, particularly, to the Newtonian potential in the
non-relativistic limit.}

 \section{Introduction}

The fascinating beauty of the  theory of the functions of complex variable reveals
itself, for example, in the harmony of the algebraic fractals on the Euclidian
plane. It makes many researches look for the analogous number systems, the elements
of which could be correlated not to the points on the plane but to the points of the
4-dimensional space-time. In case of the success of such a search, we could really
trust the famous Pythagoras saying 'all the existing is number'. On this way, the
interesting results were obtained for quaternions [1], biquaternions [2-4], octaves
[5] and so forth. Nevertheless, none of these number system theories can be compared
even to the theory of the relatively simple 2-component complex numbers. The main
reason for this seems to be the lack of the commutativity (and sometimes even of the
associativity) of the multiplication in these algebras. Although the authors of this
paper realize the conceptual bases of all the variety of algebras, the commutativity
of the multiplication is the integral property of all the principal number systems
that contain natural, integer, rational, real and complex numbers. Finally, the
commutativity and the associativity of the multiplication are among the axioms of
arithmetic which presents the foundation of mathematics, and it would be strange if
the algebraic system which is the most natural for our real world does not
correspond to the rules of regular counting.

One of the systems free from this drawback is the algebra of the commutative and
associative hyper complex numbers, related to the direct sum of the four real
algebras, which will be denoted as $H_4$. The algebra of these numbers is isomorphic
to the algebra of the 4-dimensional square real diagonal matrices, and the
corresponding space is a linear Finsler space with the Berwald-Moor metric (the last
fact was proved by the authors in [6]). It should be mentioned that Finsler space
with the Berwald-Moor metric has been known and partially investigated for a long
time [7--8].

One of the main properties of this space is the existence of such a range of the
parameters that the 3-dimensional distances (from the point of view of the observer
who uses the radar method to measure them [9]) correspond to the positively defined
metric function the limit of which is the quadratic form [10]. In other words, the
3-dimensional world observed by an "$H_4$ inhabitant" is Euclidian within certain
accuracy. Moreover, when one passes to the relativistic velocities, the
4-dimensional intervals between the $H_4$ events present the Minkowski space
correlations [11]. All this makes possible to suggest that the $H_4$ space and the
corresponding Finsler geometry can be used as a mathematical model of the real
space-time, and maybe this model would be even more productive than the pseudo
Riemannian constructions prevailing in Physics now.

\mes

Any hyper complex algebra is completely defined by the multiplication rule for the
elements of a certain fixed basis. In the H4 number system there is a special --
isotropic -- basis $e_1, e_2, e_3, e_4$, such that
 \begin{equation}\label{gp1}
 e_ie_j=p^k_{ij}e_k \, \qquad  p^k_{ij} = \left\{  \begin{array}{l}
                                                    1\, , \hbox{\q if~} \; i=j=k \, , \\
                                                    0\, , \hbox{\q else} \, .
                                                  \end{array}
   \right.
 \end{equation}
 Any analytical function in this basis can be given as
 \begin{equation}\label{gp2}
 F(X) = f^1(\xi^1)e_1 + f^2(\xi^2)e_2 + f^3(\xi^3)e_3 + f^4(\xi^4)e_4 \, ,
 \end{equation}
where
 \begin{equation}\label{gp3}
 H_4 \ni X = \xi^1e_1 + \xi^2e_2 + \xi^3e_3 + \xi^4e_4 \, ,
 \end{equation}
and $f^i$ are four arbitrary smooth functions of a single real variable.

In $H_4$ there is one more -- orthogonal -- selected basis $1, \, j, \, k, \, jk$,
which is related to the isotropic basis by the following formulas
 \begin{equation}\label{gp4}
 \left.  \begin{array}{l}
          1 = e_1 + e_2 + e_3 + e_4 \, , \\[2pt]
          j = e_1 + e_2 - e_3 - e_4 \, , \\[2pt]
          k = e_1 - e_2 + e_3 - e_4 \, , \\[2pt]
          jk = e_1 - e_2 - e_3 + e_4 \, ,
        \end{array}
  \right\}
 \end{equation}
where {\it 1} is the unity of algebra, and the corresponding component of the
analytical function of the $H_4$ variable is defined by the formula
 \begin{equation}\label{gp5}
 u = \frac{1}{4} \left[  f^1(\xi^1) + f^2(\xi^2) + f^3(\xi^3) + f^4(\xi^4)  \right] \, .
 \end{equation}

If $X$ is a radius vector, then the coordinate space  $\xi^1, \, \xi^2, \, \xi^3, \,
\xi^4$  is a Berwald-Moor space with the length element
 \begin{equation}\label{gp6}
 ds = \sqrt[4]{d\xi^1d\xi^2d\xi^3d\xi^4} \equiv
 \sqrt[4]{g_{ijkl}d\xi^id\xi^jd\xi^kd\xi^l} \, ,
 \end{equation}
where
 \begin{equation} \label{gp7}
 g_{ijkl} = \left\{
 \begin{array}{l}
  \frac{1}{4!} \, ,  {~~(i\ne j\ne k\ne l)}   , \\[9pt]
 \; 0\, \, ,  {~~(else)}  .
 \end{array} \right.
 \end{equation}
For this geometry the tangent indicatrix equation is
 \begin{equation}\label{gp8}
 g^{ijkl}p_ip_jp_kp_l - 1 = 0 \, ,
 \end{equation}
where
 \begin{equation}\label{gp9}
 g^{ijkl} = \left\{  \begin{array}{l}
 \displaystyle\frac{4^4}{4!} \, ,  {~~(i\ne j\ne k\ne l)}  , \\[9pt]
 \; 0\, \, , \hbox{~~(else)} ,
 \end{array} \right.
 \end{equation}
 \begin{equation}\label{gp10}
 p_i = \frac{g_{ijkl}d\xi^jd\xi^kd\xi^l}{\left(g_{mrst}d\xi^md\xi^rd\xi^sd\xi^t
 \right)^{3/4}} \,
 \end{equation}
are the components of the generalized momentum or generalized momenta.

If we have tensors  $p^k_{ij}$, $g_{ijkl}$, $g^{ijkl}$  and vector fields of the
analytical functions $F_{(A)}(X)$ of the $H_4$ variables, we could construct the
metric tensors in the 4-dimensional space-time in many ways. For example,
 \begin{equation}\label{gp11}
 g_{ij}(\xi) = g_{ijkl}f^k_{(1)}f^l_{(2)} \, ,
 \end{equation}
Now one can investigate the obtained Riemannian geometry. The main drawback of this
approach is the variety of the ways to construct it.

 It is known [12] that if the tangent indicatrix equation is defined as
 \begin{equation}\label{gp12}
 \Phi(p;\xi) = 0 \, ,
 \end{equation}
then the geodesics will be the solutions of the canonical system of differential
equations
 \begin{equation}\label{gp13}
 \dot{\xi}^i = \frac{\partial \Phi}{\partial p_i} \cdot \lambda(p;\xi) \, , \qquad \dot{p}_i
 = - \frac{\partial \Phi}{\partial \xi^i} \cdot \lambda(p;\xi) \, ,
 \end{equation}
 $\lambda (p;\xi )\neq 0$  is an arbitrary smooth function, and a dot above $\xi^i$ and $p_i$ means the derivation by the evolution parameter, $\tau$.

\section{Construction of the metric function\\ of the pseudo Riemannian space}

Let us regard a space which is conformally connected to the $H_4$ space, that is to
the space with the length element
 \begin{equation}\label{gp14}
 ds' = \kappa(\xi)\cdot \sqrt[4]{g_{ijkl}d\xi^id\xi^jd\xi^kd\xi^l} \, ,
 \end{equation}
where $\kappa(\xi) > 0$ is a scalar function which is a contraction-extension
coefficient depending on the point.

Let there be a normal congruence of geodesics (world lines). Then there is a scalar
function $S(\xi)$ (see, e.g. [12]) such that its level hyper surfaces are
transversal to this normal congruence of the world lines and this function is a
solution of the equation
 \begin{equation}\label{gp15}
 g^{ijkl}\frac{\partial S}{\partial \xi^i}\frac{\partial S}{\partial
 \xi^j}\frac{\partial S}{\partial \xi^k}\frac{\partial S}{\partial
 \xi^l} = \kappa(\xi)^4  \, ,
 \end{equation}
while the generalized momenta along this congruence of the world lines are related
to $S(\xi)$ by
 \begin{equation}\label{gp16}
 p_i = \frac{\partial S}{\partial \xi^i} \, ,
 \end{equation}
The equations for the world lines obtain the form
 \begin{equation}\label{gp17}
 \dot{\xi}^i = g^{ijkl}\frac{\partial S}{\partial
 \xi^j}\frac{\partial S}{\partial \xi^k}\frac{\partial S}{\partial
 \xi^l} \cdot \lambda(\xi) \, ,
 \end{equation}
were  $\lambda (\xi )\neq 0$.

In Physics the function $S(\xi)$ is called "action as a function of coordinates"\,
and (\ref{gp15}) is known as the Hamilton-Jacoby equation. In \cite{10} the function
$S(\xi)$ was called the \emph{World function}.

If there is a congruence of the world lines, then the evolution of every point in
space is known, particularly, the velocity field is known, but the energy
characteristics of the material objects (observers) corresponding to a given world
line are not known. The knowledge of the World function $S(\xi)$ makes it possible
to calculate the generalized momenta $p_i$, corresponding to the energy
characteristics, and the invariant energy characteristic, $\kappa(\xi)$, which has
also the meaning of the local contraction-extension coefficient of the plane $H_4$
space.

So, if our world view is the classical mechanics, then any pair out of the three:
World function, congruence of the world lines, Finsler geometry - gives us the
complete knowledge of the World.

Let us construct a twice contravariant tensor {\it g}{\it ij}{\it (?)} in the
following way:
 \begin{equation}\label{gp18}
 g^{ij}(\xi) = \frac{1}{\kappa(\xi)^4} \cdot g^{ijkl}\frac{\partial
 S}{\partial \xi^k}\frac{\partial S}{\partial \xi^l} \, .
 \end{equation}
Since
 \begin{equation}\label{gp19}
 det(g^{ij}(\xi))= - \frac{4^4}{3^3 \kappa(\xi)^8} \neq 0 \, ,
 \end{equation}
then everywhere where the geometry (\ref{gp14}) is defined, one can construct a
tensor $g_{ij}(\xi)$ such that
 \begin{equation}\label{gp20}
 g^{ik}(\xi)g_{kj}(\xi)=\delta^i_j  \, ,
 \end{equation}
 \begin{equation}\label{gp21}
  g_{ij}(\xi)   = 4\cdot \left(
 \begin{array}{cccc}
  -2\left(\frac{\partial S}{\partial \xi^1}\right)^2 & \frac{\partial S}{\partial \xi^1}\frac{\partial S}{\partial \xi^2} & \frac{\partial S}{\partial \xi^1}\frac{\partial S}{\partial \xi^3} & \frac{\partial S}{\partial \xi^1}\frac{\partial S}{\partial \xi^4}
  \\[9pt]
  \frac{\partial S}{\partial \xi^1}\frac{\partial S}{\partial \xi^2} & -2\left(\frac{\partial S}{\partial \xi^2}\right)^2  & \frac{\partial S}{\partial \xi^2}\frac{\partial S}{\partial \xi^3} & \frac{\partial S}{\partial \xi^2}\frac{\partial S}{\partial \xi^4} \\[9pt]
  \frac{\partial S}{\partial \xi^1}\frac{\partial S}{\partial \xi^3} & \frac{\partial S}{\partial \xi^2}\frac{\partial S}{\partial \xi^3} & -2\left(\frac{\partial S}{\partial \xi^3}\right)^2  & \frac{\partial S}{\partial \xi^3}\frac{\partial S}{\partial \xi^4} \\[9pt]
  \frac{\partial S}{\partial \xi^1}\frac{\partial S}{\partial \xi^4} & \frac{\partial S}{\partial \xi^2}\frac{\partial S}{\partial \xi^4} & \frac{\partial S}{\partial \xi^3}\frac{\partial S}{\partial \xi^4} & -2\left(\frac{\partial S}{\partial \xi^4}\right)^2
 \end{array}
 \right) .
 \end{equation}

No doubt that in the same coordinate space  $\xi ^{1} ,\xi ^{2} ,\xi ^{3} ,\xi ^{4}$
such tensor $g_{ij}(\xi)$ defines a Riemannian or pseudo Riemannian geometry with
the length element
 \begin{equation}\label{gp22}
 ds'' = \sqrt{g_{ij}(\xi)d\xi^id\xi^j} \, .
 \end{equation}

The construction of tensor $g_{ij}(\xi)$ leads directly to the conclusion: the
change of geometry (\ref{gp14}) to the geometry (\ref{gp22}) does not lead to the
change of the initial congruence of the world lines and corresponding World function
$S(\xi)$.

Therefore, in our concept one and the same World, i.e. the pair \{World function;
congruence of the world lines\}, corresponds to a whole class of related but
qualitatively different Finsler geometries.

\section{Analyticity condition and the Minkowski space}

Let the World function $S(\xi)$ be the (unity) component of an analytical function
of the $H_4$ variable in the orthogonal basis (\ref{gp4}), that is
 \begin{equation}\label{gp23}
 S(\xi) = \frac{1}{4} \left[  f^1(\xi^1) + f^2(\xi^2) + f^3(\xi^3) + f^4(\xi^4)  \right] \, .
 \end{equation}
Then
 \begin{equation}\label{gp24}
 g^{ijkl}\frac{\partial S}{\partial \xi^i}\frac{\partial S}{\partial
 \xi^j}\frac{\partial S}{\partial \xi^k}\frac{\partial S}{\partial
 \xi^l} = \frac{\partial f^1(\xi^1)}{\partial \xi^1}\frac{\partial
 f^2(\xi^2)}{\partial \xi^2}\frac{\partial f^3(\xi^3)}{\partial
 \xi^3}\frac{\partial f^4(\xi^4)}{\partial \xi^4} = \kappa(\xi)^4 > 0  \, ,
 \end{equation}
and this leads to the limitation on the functions, {\it f}{\it i}:
 \begin{equation}\label{gp25}
 \frac{\partial f^1(\xi^1)}{\partial \xi^1}\frac{\partial
 f^2(\xi^2)}{\partial \xi^2}\frac{\partial f^3(\xi^3)}{\partial
 \xi^3}\frac{\partial f^4(\xi^4)}{\partial \xi^4} > 0 \, .
 \end{equation}

It follows from (\ref{gp24}) that the space with the length element (\ref{gp14}) can
be obtained from the space with the length element (\ref{gp6}) with the help of the
conformal transformation, which means that the condition of the analyticity of the
World function can be treated in a sense as the condition of the conformal symmetry.

Let us construct tensor $g_{ij}(\xi)$ following the algorithm developed in the
previous section. It turns out that in a region where functions $f^i$  have no
singularities there will always be such a coordinate system $x^0, \, x^1, \, x^2, \,
x^3$ in which the length element $ds''$ has a form
 \begin{equation}\label{gp26}
 ds'' = \sqrt{(x^0)^2 - (x^1)^2 - (x^3)^2 - (x^3)^2} \, .
 \end{equation}

Let us express the coordinates $x^0,\, x^1,\, x^2,\, x^3$ in terms of the initial
coordinates  $\xi ^{1} ,\xi ^{2} ,\xi ^{3} ,\xi ^{4} $:
\begin{equation}\label{gp27}
\left.
\begin{array}{l}
  x^0 = \displaystyle\frac{\;\, 1 \;}{4}       \left(  f^1(\xi^1) + f^2(\xi^2) + f^3(\xi^3) +
f^4(\xi^4)  \right) \, ,\\[12pt]
  x^1 = \displaystyle\frac{\sqrt{3}}{4}\left(  f^1(\xi^1) + f^2(\xi^2) - f^3(\xi^3) -
f^4(\xi^4)  \right)  \, ,\\[12pt]
  x^2 = \displaystyle\frac{\sqrt{3}}{4}\left(  f^1(\xi^1) - f^2(\xi^2) + f^3(\xi^3) -
f^4(\xi^4)  \right) \, ,\\[12pt]
  x^3 =\displaystyle \frac{\sqrt{3}}{4}\left(  f^1(\xi^1) - f^2(\xi^2) - f^3(\xi^3) +
f^4(\xi^4)  \right) \, .
\end{array}
\right\}
\end{equation}

Therefore, to obtain the non-trivial curving of the space-time one should use the World functions with the broken conformal symmetry.

\section{Newtonian  potential}

Let us show that there are World functions that lead to the non-trivial pseudo
Riemannian 4-dimensional spaces. Let us regard a function
 \begin{equation}\label{gp28}
 S(\xi) = \frac{1}{4}\left( \xi^1 + \xi^2 + \xi^3 + \xi^4 \right) +
 \alpha\cdot\psi(\varrho) \, ,
 \end{equation}
where $\alpha$ is the parameter characterizing the break of the analyticity of the
World function (the break of the conformal symmetry in the $H_4$ space),  $\psi$ is
an arbitrary function of a single argument
 \begin{equation}\label{gp29}
 \varrho = \sqrt{(y^1)^2+(y^2)^2+(y^3)^2} \, ,
 \end{equation}
and $y^0,\, y^1,\, y^2,\, y^3$ are the coordinates in the orthogonal basis $1, j, k,
jk$:
 \begin{equation}\label{gp30}
 \left.
 \begin{array}{c}
 y^0 = \displaystyle \frac{1}{4}(\xi^1+\xi^2+\xi^3+\xi^4) \, , \\[12pt]
 y^1 = \displaystyle \frac{1}{4}(\xi^1+\xi^2-\xi^3-\xi^4) \, , \\[12pt]
 y^2 = \displaystyle \frac{1}{4}(\xi^1-\xi^2+\xi^3-\xi^4) \, , \\[12pt]
 y^3 = \displaystyle \frac{1}{4}(\xi^1-\xi^2-\xi^3+\xi^4) \, .
 \end{array}
 \right\} \,
 \end{equation}
Then the derivatives of the World functions over the coordinates $\xi^i$  can be
expressed in the following way:
\begin{equation}\label{gp31}
\left.
\begin{array}{c}
\displaystyle\frac{\partial S}{\partial \xi^1} =
\displaystyle \frac{1}{4}\left[1+\frac{\alpha}{\varrho}\frac{d\psi}{d\varrho}\left( y^1+y^2+y^3  \right)\right] \, , \\[15pt]
\displaystyle\frac{\partial S}{\partial \xi^2} =
\displaystyle \frac{1}{4}\left[1+\frac{\alpha}{\varrho}\frac{d\psi}{d\varrho}\left( y^1-y^2-y^3  \right)\right] \, , \\[15pt]
\displaystyle\frac{\partial S}{\partial \xi^3} =
\displaystyle \frac{1}{4}\left[1+\frac{\alpha}{\varrho}\frac{d\psi}{d\varrho}\left( -y^1+y^2-y^3  \right)\right] \, , \\[15pt]
\displaystyle\frac{\partial S}{\partial \xi^4} =
\displaystyle\frac{1}{4}\left[1+\frac{\alpha}{\varrho}\frac{d\psi}{d\varrho}\left(
-y^1-y^2+y^3  \right)\right] \, .
\end{array}
\right\} \,
\end{equation}

Let us calculate the components of the metric tensor in coordinates $y^0,\, y^1,\,
y^2,\, y^3$ using the invariance of the square of the length element
 \begin{equation}\label{gp32}
 g_{ij}(\xi) d\xi^id\xi^j = \tilde{g}_{ij}(y)dy^idy^j
 \end{equation}
Grouping the terms, one gets
 \begin{equation}\label{gp33}
 \tilde{g}_{00}= 1 - 3\alpha^2\left( \frac{d\psi}{d\varrho} \right)^2
 \, , \qquad \tilde{g}_{\beta\beta_-}=-3\left\{1 + \alpha^2\left(
 \frac{d\psi}{d\varrho} \right)^2\left[ 1 -
 \frac{4(y^\alpha)^2}{3\rho^2}\right]\right\} \, ,
 \end{equation}
 \begin{equation}\label{gp34}
 2\tilde{g_{0\beta}} = - 4 \left[
 \alpha\frac{d\psi}{d\varrho}\frac{\;\; y^\beta}{\varrho} +
 3\alpha^2\left(\frac{d\psi}{d\varrho}\right)^2\cdot\frac{y^1y^2y^3}{y^\beta\varrho^2}
 \right] \,  ,
 \end{equation}
 \begin{equation}\label{gp35}
 2\tilde{g}_{\beta\gamma} = - 4 \left[ 3
 \alpha\frac{d\psi}{d\varrho}\frac{\;\, y^\delta}{\varrho} +
 \alpha^2\left(\frac{d\psi}{d\varrho}\right)^2\cdot\frac{y^\beta
 y^\gamma}{\varrho^2} \right] \, ,
 \end{equation}
where $\beta,\, \gamma,\, \delta,\,  = 1, 2, 3$; $\beta\equiv\beta_-$ but no
summation is performed here; in the last formula all the indices $\beta,\, \gamma,\,
\delta\,$ are different.

 If $\alpha = 0$, then
 \begin{equation}\label{gp36}
 (\tilde{g}_{ij}) = diag(1,-3,-3,-3) \, .
 \end{equation}
This means that the real physical coordinates $x^0,\, x^1,\, x^2,\, x^3$ of the
space-time are expressed by the coordinates $y^0,\, y^1,\, y^2,\, y^3$ in the
following way
 \begin{equation}\label{gp37}
 x^0 = y^0 \, , \qquad x^\beta = \sqrt{3}\cdot y^\beta \, .
 \end{equation}

Let us pass to the physical coordinates $x^0,\, x^1,\, x^2,\, x^3$:
 \begin{equation}\label{gp38}
 \tilde{g}_{ij}(y)dy^idy^j = \bar{g}_{ij}(x)dx^idx^j \, ,
 \end{equation}
where
 \begin{equation}\label{gp39}
 \bar{g}_{00} =   \tilde{g}_{00} \, , \qquad \bar{g}_{0\beta} =
 \frac{1}{\sqrt{3}}\cdot \tilde{g}_{0\beta} \, , \qquad \bar{g}_{\beta\gamma} =
 \frac{1}{3}\cdot \tilde{g}_{\beta\gamma} \,.
 \end{equation}
 Let us denote
 \begin{equation}\label{gp40}
 r = \sqrt{(x^1)^2+(x^2)^2+(x^3)^2} \equiv \sqrt{3}\cdot\varrho \, ,
 \end{equation}
Then
 \begin{equation}\label{gp41}
 \bar{g}_{00}= 1 - 9\alpha^2\left( \frac{d\psi}{dr} \right)^2 \, ,
 \qquad \bar{g}_{\beta\beta_-}=-\left\{1 + 3\alpha^2\left(
 \frac{d\psi}{dr} \right)^2\left[ 1 -
 \frac{4(x^\alpha)^2}{3r^2}\right]\right\} \, ,
 \end{equation}
 \begin{equation}\label{gp42}
 2\bar{g_{0\beta}} = - 4 \left[ \alpha\frac{d\psi}{dr}\frac{\;\;
 x^\beta}{r} +
 3\sqrt{3}\alpha^2\left(\frac{d\psi}{dr}\right)^2\cdot\frac{x^1x^2x^3}{x^\beta
 r^2} \right] \,  ,
 \end{equation}
 \begin{equation}\label{gp43}
 2\bar{g}_{\beta\gamma} = - 4 \left[ \sqrt{3}
 \alpha\frac{d\psi}{dr}\frac{\;\, x^\delta}{r} +
 \alpha^2\left(\frac{d\psi}{dr}\right)^2\cdot\frac{x^\beta x^\gamma}{r^2} \right] \, .
 \end{equation}

The metric tensor  $\bar{g}_{ij} (x)=\bar{g}_{ij} (x^{1} ,x^{2} ,x^{3} )$  depends
only on the space coordinates $x^1,x^2,x^3$, and this corresponds to the stationary
gravitational field, stationary Universe. The probe particle of mass $m$ moves along
the geodesic of the pseudo Riemannian space with metric tensor  $\bar{g}_{ij} (x^{1}
,x^{2} ,x^{3} )$.

Let a particle move in a fixed frame and have velocity much less than the light
velocity, $c$:
 \begin{equation}\label{gp44}
 \frac{dx^\beta}{dt} = v^\beta \, , \qquad   |v^\beta| \ll c \, ,
 \end{equation}
The gravitational fields are weak, that is the condition  $|v^{\beta } |<<1$ remains
valid for all the time of the particle motion. Let us obtain the Lagrange function,
{\it L}, to describe such non-relativistic motion of the probe particle in the weak
gravity field. To do this, develop the right hand side of the expression
 \begin{equation}\label{gp45}
 L = -mc\cdot \frac{\sqrt{\bar{g}_{ij}(x^1,x^2,x^3)dx^idx^j}}{dt}
 \end{equation}
Within the accuracy of $\left(\frac{v}{c}\right)^2$
\begin{equation}\label{gp46}
L = -mc^2 \sqrt{\bar{g}_{00}} \cdot \sqrt{1+ \frac{1}{\bar{g}_{00}}\left(
2\bar{g}_{0\beta}\frac{v^\beta}{c} + \bar{g}_{\beta\gamma}\frac{v^\beta
v^\gamma}{c^2} \right)} \, ,
\end{equation}
\begin{equation}\label{gp47}
L \simeq -mc^2 \sqrt{\bar{g}_{00}} \cdot \left\{1+ \frac{1}{2\bar{g}_{00}}\left(
2\bar{g}_{0\beta}\frac{v^\beta}{c} + \bar{g}_{\beta\gamma}\frac{v^\beta
v^\gamma}{c^2} \right) - \frac{1}{8\bar{g}^2_{00}}\left(
2\bar{g}_{0\beta}\frac{v^\beta}{c} \right)^2\right\} \, .
\end{equation}
Opening the brackets in the right hand side, we get an additive term which is the
full time derivative of a certain function {\it f(r)}, it depends linearly on the
velocity components and, thus, it can be omitted. Leaving the same designation for
the Lagrange function, we get
\begin{equation}\label{gp48}
L \simeq -mc^2 \sqrt{\bar{g}_{00}} \cdot \left\{1+ \frac{1}{2\bar{g}_{00}}\cdot
\bar{g}_{\beta\gamma}\frac{v^\beta v^\gamma}{c^2} - \frac{1}{8\bar{g}^2_{00}}\left(
2\bar{g}_{0\beta}\frac{v^\beta}{c} \right)^2\right\} \, .
\end{equation}

Our goal is the Lagrange function of the form
 \begin{equation}\label{gp49}
 L = \frac{m\vec{v}^2}{2} - U(\vec{x}) \, ,
 \end{equation}
where  $U(\vec{x})$  is the potential energy of the probe particle,  $\vec{x}\equiv
(x^{1} ,x^{2} ,x^{3} ),~ \vec{v}\equiv (v^{1} ,v^{2} ,v^{3} ),~ r^{2} =
\vec{x}^{\,2}$,  $\vec{v}^{\,2} =(v^{1} )^{2} +(v^{2} )^{2} +(v^{3} )^{2} \equiv
v^{2} $. To reach it we have to make some assumptions about the correlation between
the parameter, $\alpha$ and light velocity:
 \begin{equation}\label{gp50}
 \alpha = \frac{\nu}{c} \, , \quad \hbox{when} \quad c\rightarrow
 \infty \quad \alpha\rightarrow 0 \, .
 \end{equation}
Besides, let $\alpha$ be of the same order (or smaller) with the relation
$\left|\displaystyle\frac{v}{c}\right|$. Then leaving only the terms that don't
disappear at  $c\to \infty $  in the (\ref{gp48}), one gets
 \begin{equation}\label{gp51}
 L \simeq -mc^2 + mc^2 \frac{9}{2}\frac{\nu^2}{c^2} \left(
 \frac{d\psi}{dr} \right)^2 + m\cdot \frac{v^1v^1+v^2v^2+v^3v^3}{2}\, .
 \end{equation}
Since $(-mc^2)$ is a full time derivative of function $(-mc^2\cdot t)$, we omit it
and get
 \begin{equation}\label{gp52}
 L \simeq  \frac{m\vec{v}^2}{2} + \frac{9m\nu^2}{2} \left(
 \frac{d\psi}{dr} \right)^2 \, .
 \end{equation}

Let a mass $M$ be motionless in the frame origin, and then the potential energy of
the probe particle with mass $m$ located at $x^1,x^2,x^3$ is equal to
 \begin{equation}\label{gp53}
 U(r) = - \gamma \frac{mM}{r} \, ,
 \end{equation}
where $\gamma$ is the gravitational constant. Comparing (\ref{gp49}) and
(\ref{gp52}), we get the equation for $\psi(r)$:
\begin{equation}\label{gp54}
\frac{9m\nu^2}{2} \left(\frac{d\psi}{dr} \right)^2 = \gamma \frac{mM}{r} \qquad
\Rightarrow \qquad  \frac{d\psi}{dr} = \pm \frac{\sqrt{2\gamma
M}}{3\nu}\frac{1}{r^{1/2}}   \, .
\end{equation}
Therefore,
\begin{equation}\label{gp55}
\psi(r) = \pm \frac{2\sqrt{2\gamma M}}{3\nu}\cdot r^{1/2} + \psi_0  \qq (\psi_0 =
const).
\end{equation}

Finally, the World function is equal to
\begin{equation}\label{gp56}
S = x^0 \pm \frac{2\sqrt{2\gamma M}}{3c}\cdot r^{1/2} + C_0 \qq (C_0 = const),
\end{equation}
When it performs a conformal transformation of the length element of the plane
Berwald-Moor space, it induces a pseudo Riemannian geometry in the Minkowski space.
For a non-relativistic probe particle of mass {\it m}, this geometry gives the
motion equations for the Kepler problem for the point mass {\it M} located in the
origin of the space frame.

 The more complicated World function, maybe also leading to the stationary Universe, has the form
\begin{equation}\label{gp57}
S(\xi) = \frac{1}{4}\left( \xi^1 + \xi^2 + \xi^3 + \xi^4 \right)\left[
1+\alpha_1\cdot\psi_1(\varrho) \right] + \alpha_2\cdot\psi_2(\varrho) \, ,
\end{equation}
where $\alpha_A$ are the parameters of the analyticity break of the World function
(parameters of the conformal symmetry break in the $H_4$ space), $\psi_A$ are the
arbitrary functions of single argument $\varrho$ (\ref{gp29}), (\ref{gp30}).

 \section*{Conclusion}

The results obtained in this paper point at the deep correlation between the
Einstein geometries and Finsler spaces with Berwald-Moor metric. We managed to find
the concrete Finsler space with the Berwald-Moor metric which in the limit appeared
to be related to the curved pseudo Riemannian space with the Newtonian gravitational
potential. This fact points at the principal possibility to built more interesting
constructions, particularly, such Finsler spaces whose limit cases would be the
known relativistic solutions.

\end{document}